\documentclass[aps,prd,preprint,superscriptaddress,tightenlines,nofootinbib,showpacs]{revtex4}

\usepackage{graphicx}
\usepackage{dcolumn}
\usepackage{bm}
\usepackage{epsfig}

\begin{document}

\preprint{CLNS 02/1812}       
\preprint{WSU-HEP-0210}         

\title{A Method for Measuring Emittance in $e^{+} e^{-}$ Colliding Beams}


\author{D. Cinabro}
\affiliation{Wayne State University, Detroit, MI 48202, USA}
\author{K. Korbiak\footnote{Present address University of Michigan, Ann Arbor, MI 48109, USA}}
\affiliation{Wayne State University, Detroit, MI 48202, USA}
\author{M. Billing}
\affiliation{Cornell University, Ithaca, NY 14853, USA}
\author{N. Mistry}
\affiliation{Cornell University, Ithaca, NY 14853, USA}
\author{D. Rice}
\affiliation{Cornell University, Ithaca, NY 14853, USA}
\author{D. Rubin}
\affiliation{Cornell University, Ithaca, NY 14853, USA}

\date{4 December 2002}

\begin{abstract} 
We have developed techniques that allow simultaneous measurement of the 
spatial size of the luminous colliding beam 
region and the angular spread of beams in collision using 
$e^{+} e^{-} \to  \mu^{+} \mu^{-}$ events.  These are
demonstrated at the CLEO 
interaction point of the Cornell Electron-Positron Storage Ring, CESR,  
taking advantage of the 
small and well understood resolution of the CLEO 
tracking system.  These measurements are then used to extract the
horizontal beta, horizontal emittance and the vertical emittance
and search for dynamic effects at CESR.
\end{abstract}

\maketitle

The measurement of the emittance of colliding beams is a difficult and 
important problem as many methods are destructive or disruptive to the beams
or involve extrapolation using a theoretical accelerator transport
matrix \cite{seeman}.  Here we present a method that relies 
on the tracking system of a high energy physics experiment
done while the beams are producing luminosity without disruption or
use of a transport matrix.  The basic technique is to select an ensemble of
$e^{+} e^{-} \to \mu^{+} \mu^{-}$ events with which we
simultaneously measure the size of the luminous region and the 
intrinsic angular spread of the outgoing particles.  At CESR and CLEO
the resolution is precise enough that we can extract the the
horizontal beta at the interaction point, $\beta_x^{\ast}$,
the horizontal emittance, $\epsilon_x$,
and, combined with our previously described technique for measuring
vertical beta at the interaction point, $\beta_y^{\ast}$~\cite{hourglass}, the
vertical emittance, $\epsilon_y$.

Assuming that dispersion and coupling are negligible, the beam parameters
$\beta_x^{\ast}$, $\beta_y^{\ast}$, $\epsilon_x$, 
and $\epsilon_y$ are related to the physical observables $\sigma$, 
the Gaussian width of the beam at the 
interaction point, and $\sigma'$, 
the Gaussian width of the angular spread of the beam at the 
interaction point, by the following equations:
\begin{eqnarray}
\label{eq:shapex} \sigma_x  & = & \sqrt{\beta_x^{\ast} \epsilon_x} \\ 
\label{eq:shapey}\sigma_y  & = & \sqrt{\beta_y^{\ast} \epsilon_y} \\
\label{eq:angx}\sigma_x' & = & \sqrt{\epsilon_x/\beta_x^{\ast}} \\
\label{eq:angy}\sigma_y' & = & \sqrt{\epsilon_y/\beta_y^{\ast}}.
\end{eqnarray}
A simultaneous measurement of both $\sigma$ and $\sigma'$ in the collision 
region would yield measurements of $\beta^{\ast}$ and $\epsilon$ via the 
obvious algebraic manipulation.

	Our previous work has shown how the size of the luminous
region can be measured using $e^+e^- \to \mu^+\mu^-$ events~\cite{hourglass}.
Briefly a fiducial box is centered on the measured center of the 
luminous region.  The average positions of tracks passing
through the box are used to measure the size and
shape of the luminous region.  Tracks that pass through the box
nearly perpendicular to one of the sides
are most useful for making a measurement.  At the CESR-CLEO
interaction point the luminous region is roughly 7~$\mu$m high
and 300 $\mu$m wide.  Thus we select more useful tracks for the 
horizontal measurement than the vertical.  We have previously 
used this technique to measure $\beta_y^{\ast}$ and observe the hourglass 
effect at the CESR-CLEO interaction point~\cite{hourglass}.

\begin{figure}
\begin{center}
\epsfig{file=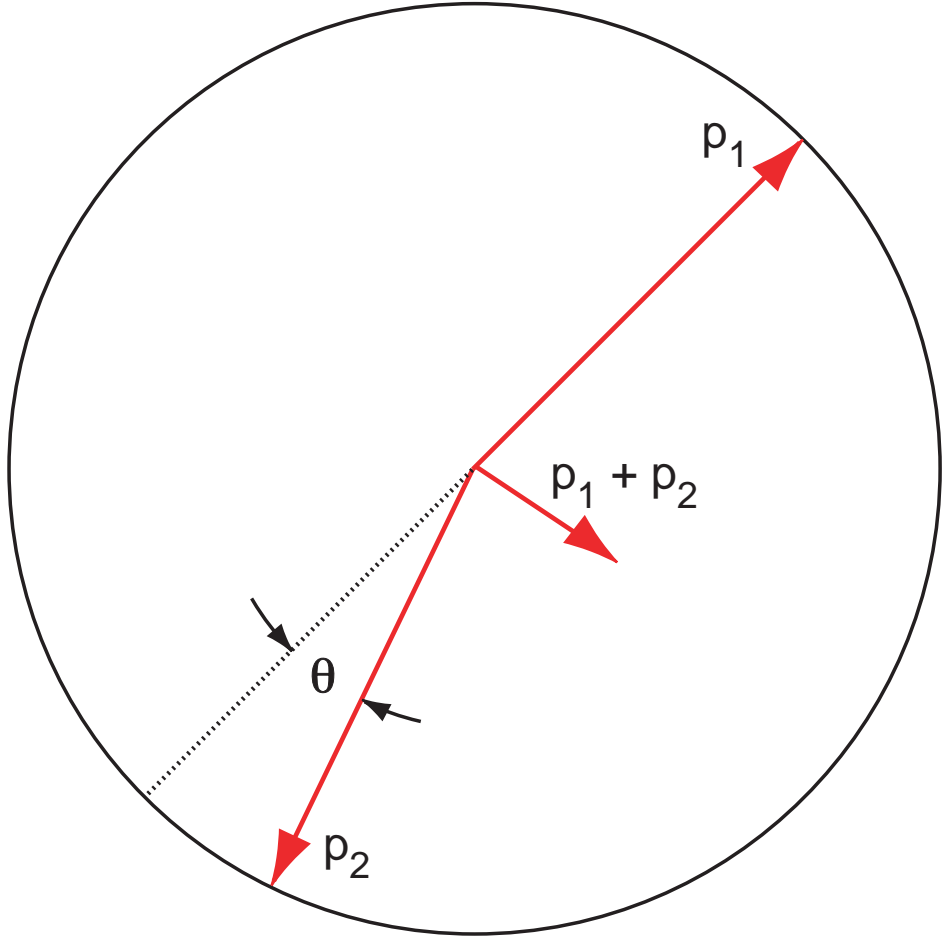,width=6.00in}
\caption{In $e^+e^- \to \mu^+\mu^-$ the momentum sum of the two tracks
is related to the angle between the two tracks. 
It gives the minor arc length which is approximately
$E_{\rm beam} \theta$.}
\label{fig:circle}
\end{center}
\end{figure}

	We measure the angle between the projected $\mu^+$ and $\mu^-$
tracks by considering the sum of the momenta of the
two tracks divided by the beam energy to obtain a vector.
If the two tracks are coming from the same point and have nearly
the same momentum as the beam
this vector is equivalent to the minor arc between the
colliding tracks, given by $E_{\rm beam} \sin \theta$, where $\theta$ is the
central angle between the colliding tracks.  This angle is
small and we approximate $\sin \theta$ by $\theta$.  We then
define the angle between the two tracks in the
horizontal projection as the horizontal momentum sum divided by
beam energy and similarly in the vertical direction.  Schematically
this is shown in Figure~\ref{fig:circle}.
 
A crossing angle between the beams causes the distribution of 
the angle to have a mean different from zero, an 
angular spread of the beam, $\sigma'$,
gives an intrinsic underlying width, and
detector momentum resolution broadens the distribution.

	CESR has been described in detail elsewhere~\cite{CESR}.
All the data used in this measurement were taken at an $e^+e^-$ 
collision energy of 10.58 GeV with bunch currents in the range of 2.5
to 7.0 mA over a four month period in late 1998 and early 1999.  The 
CLEO detector has also been described in detail elsewhere~\cite{CLEO}.
All of the data used in this measurement are taken in the CLEO~II.V
configuration which includes a silicon strip vertex detector (SVX) which
is crucial to the measurement of the size of the luminous region.
This detector consists of 
three layers of silicon wafers arrayed in an octagonal geometry around 
the interaction point.  The first measurement layer is at a radius of 
2.3 cm and the wafers are read out on both sides by strips which are 
perpendicular to each other.  The readout strips have a pitch of about 
100 $\mu$m; with charge sharing, the detector has an intrinsic
per-point resolution of better than 20$\ \mu$m both longitudinally
and in the plane transverse to the beam direction.
The rest of the tracking system consists
of a small 10 layer drift chamber, and a large 51 layer drift chamber which
has an outer radius of 1.5 m.  The tracking system is in a 1.5 Tesla magnetic
field provided by a superconducting solenoid.

	$e^+e^- \to \mu^+\mu^-$ events are easily 
selected in CLEO by choosing events with two and only two tracks each with
momentum near the beam 
energy and a small energy deposit in the electromagnetic calorimeter.  
We chose tracks with 20 or more hits in the main drift chamber, and at least 
two silicon vertex detector hits in the transverse and longitudinal views.
We require that the tracks have opposite charge and 
that those used for the measurement of the 
luminous region have at least three silicon vertex detector
hits in one of the two views.  Each point for measuring the $\sigma'$
is based 
roughly on 7000 events for $\sigma_x'$ and $\sigma_y'$ 2000 events.

	The expected beam parameters for CESR for the data discussed
in this paper are given in Table~\ref{tab:beam}.
\begin{table}
\caption{CESR beam parameters in the limit of zero bunch current.}
\begin{center}
\begin{tabular}{|c|c|}
\hline
Parameter    & Value ($\mu$m) \\ \hline
$\beta_x^\ast$    & $1.1996 \times 10^6$ \\
$\beta_y^\ast$    & 17900 \\
$\epsilon_x$ & 0.21 \\
$\epsilon_y$ & 0.0010 \\ \hline
\end{tabular}
\end{center}
\label{tab:beam}
\end{table}
All the parameters in Table~\ref{tab:beam} 
are given in the limit of zero bunch current.  
Our measurements were taken over a long time,
about four months of CESR and CLEO running, and under many different
machine conditions.  Thus we expect only rough agreement with the
parameters given in Table~\ref{tab:beam}, but we should be
sensitive to dynamic effects caused by the non-zero bunch
currents.  The horizontal $\beta^{\ast}$ and $\epsilon$ are expected to depend 
primarily on the beam current, while the vertical $\epsilon$ is 
expected to be influenced more by the measured luminosity over beam current
(the specific luminosity).  Therefore we 
present measurements in the horizontal direction as functions of bunch 
current, while measurements in the vertical are presented as functions 
of specific luminosity.  We have previously observed that $\beta_x^\ast$ 
is reduced by roughly a factor of two in colliding beam conditions
due to the dynamic beta effect~\cite{dbeta}.  Dynamic effects
are also expected to produce a larger $\epsilon_y$ and a smaller
$\beta_y^\ast$ and $\beta_x^\ast$ than given in Table~\ref{tab:beam}.

	To be able to measure the angular spread of
the colliding beams the momentum resolution must be well understood.
We study CLEO's momentum resolution with a 
simulated sample of $e^{+} e^{-} \to \mu^{+} \mu^{-}$
events with no momentum spread in the incoming beam
particles.  Recent examples show that, for high
momentum tracks, the simulation of the tracking accurately models 
the data~\cite{dstar}\cite{hourglass}. 
The resolution on the angle between the $\mu^+$ and
the $\mu^-$ depends on global properties 
of the event such as the pattern of hits
in the transverse and longitudinal views of the SVX.  More hits improves
the resolution and the $r\phi$ hits primarily effect the
resolution transverse to the beam, while $z$ hits effect
the longitudinal resolution.  Thus the data are divided according to the
SVX hit pattern.  For each sub-sample, the
resolution is  parameterized as
\begin{equation}\label{eq:res}
A + B/Nhits = \rm{Resolution}
\end{equation}
where $Nhits$ is the total number of tracking hits on
both tracks.  
We determine the width of the momentum resolution as 
a function of $Nhits$, by selecting on a range of $Nhits$ and 
performing a one dimensional fit to a 
Gaussian plus a flat background to account for events not coming from
beam collisions, mainly cosmic rays.  
To determine $A$ and $B$ for each SVX hit
pattern, we fit the two dimensional distribution of
resolution width versus $Nhits$. 
This gives us a set of $A$'s and $B$'s and combined with
the observed SVX hit pattern and number of tracking hits we determine
the resolution for each data event.
The resolutions we extract on the angle between the $\mu^+$ 
and $\mu^-$ is better than 1 mrad in both the vertical and
horizontal directions for the highest number of tracking hits
and SVX hits and up to 3 mrads for the worst case.

	We then fit the two dimensional data distributions
of angular width versus number of tracking hits for each SVX hit pattern
with a width that is quadratic sum of the resolution piece determined from the 
simulation as described above and an extra contribution due to the
underlying angular width of the colliding beams.
We note that the means of the angular distributions are consistent
with the known crossing angle, 2.0 mrads in the horizontal and none
in the vertical, with an accuracy of better than a 0.01 mrads. 
The underlying width is then the weighted average of the fitted underlying
widths over the SVX hit patterns.
Systematic uncertainties on the extracted angular width of the
colliding beams are applied by doing the data fits again with the values
for the $A$ and $B$ parameters of the resolutions varied up
or down in concert by one standard deviation based on the
results from the simulation.  Typically this
represents a variation of 0.2--0.4 mrad.  We also repeat the analysis
using only the events with the largest number of tracking hits with
the best resolution and see negligible changes.
 
	We are not sensitive to the
vertical width of the luminous region.  Our resolution on the size of the
luminous region is about $30 \mu$m while our earlier observation showed that
the underlying width is about $7 \mu$m.  We take the observed vertical width
to be the resolution on the horizontal width, and this agrees well
with the prediction of the simulation.  For this reason, dynamic
measurements of $\beta_y^{\ast}$ cannot be extracted.  A constant measured
value, $\beta_y^{\ast} = (17910 \pm 170)\mu$m~\cite{hourglass}, is used in this
analysis to extract $\epsilon_y$.  

\begin{figure}
\begin{center}
\epsfig{file=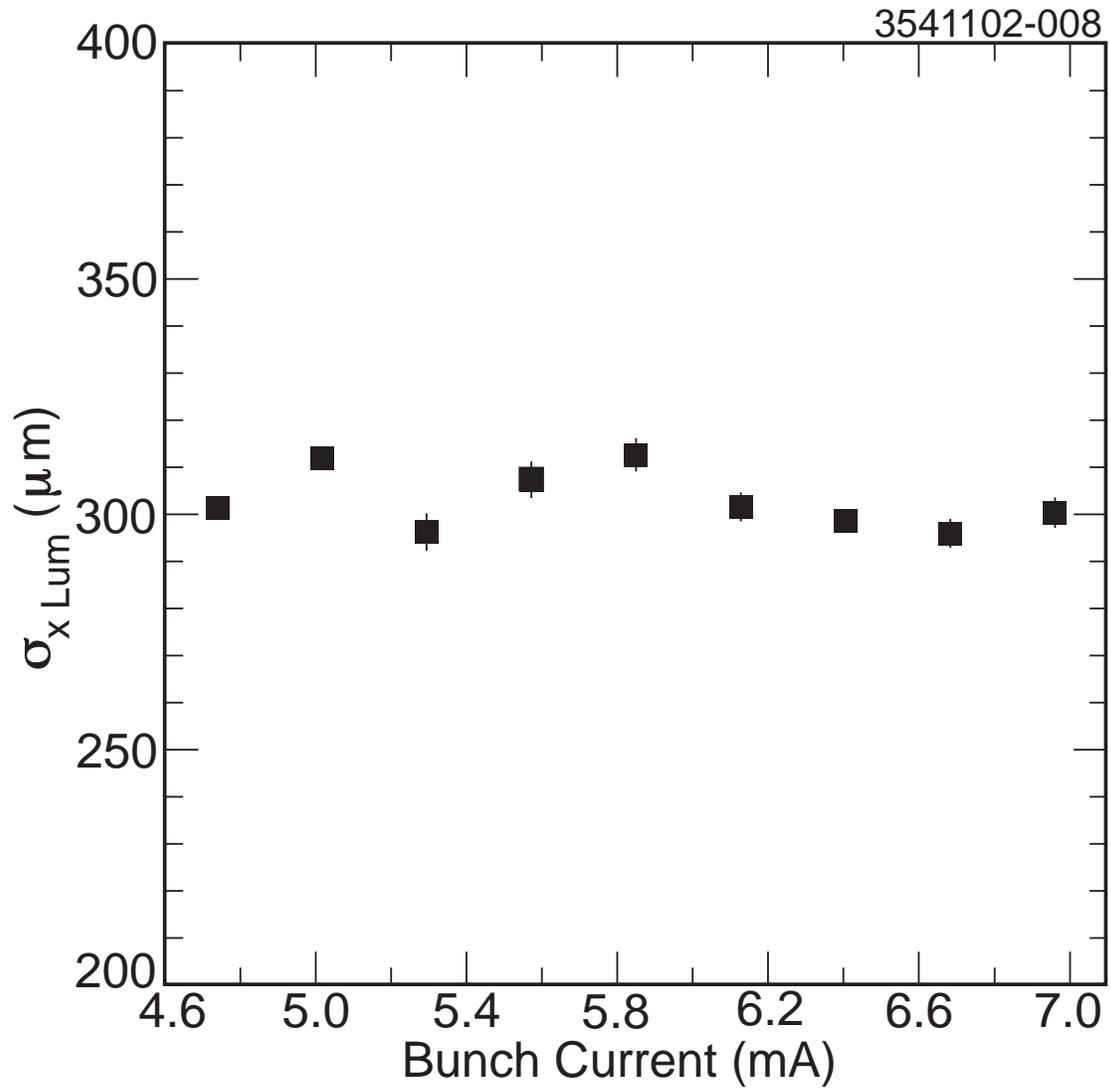,width=6.25in}
\caption{\label{fig:xsigVbmI}The Gaussian horizontal width of the 
luminous region as a function of bunch current.}
\vspace*{2.75in}
\end{center}
\end{figure}

\begin{figure}
\begin{center}
\epsfig{file=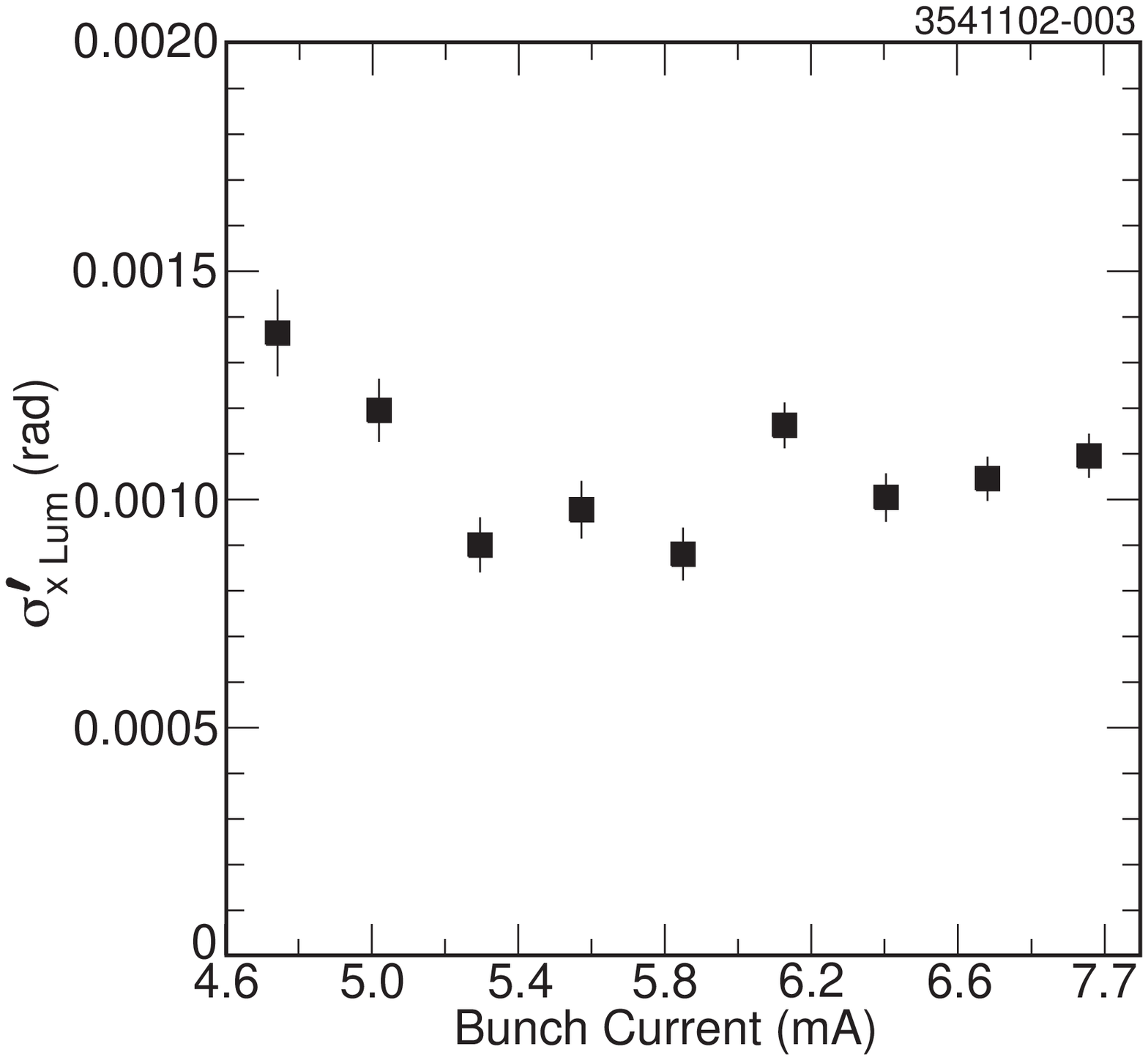,width=6.25in}
\caption{\label{fig:xsigpVbmI}The horizontal width of the angular 
distribution of the luminous region as a function of bunch current.}
\vspace*{2.75in}
\end{center}
\end{figure}

\begin{figure}
\begin{center}
\epsfig{file=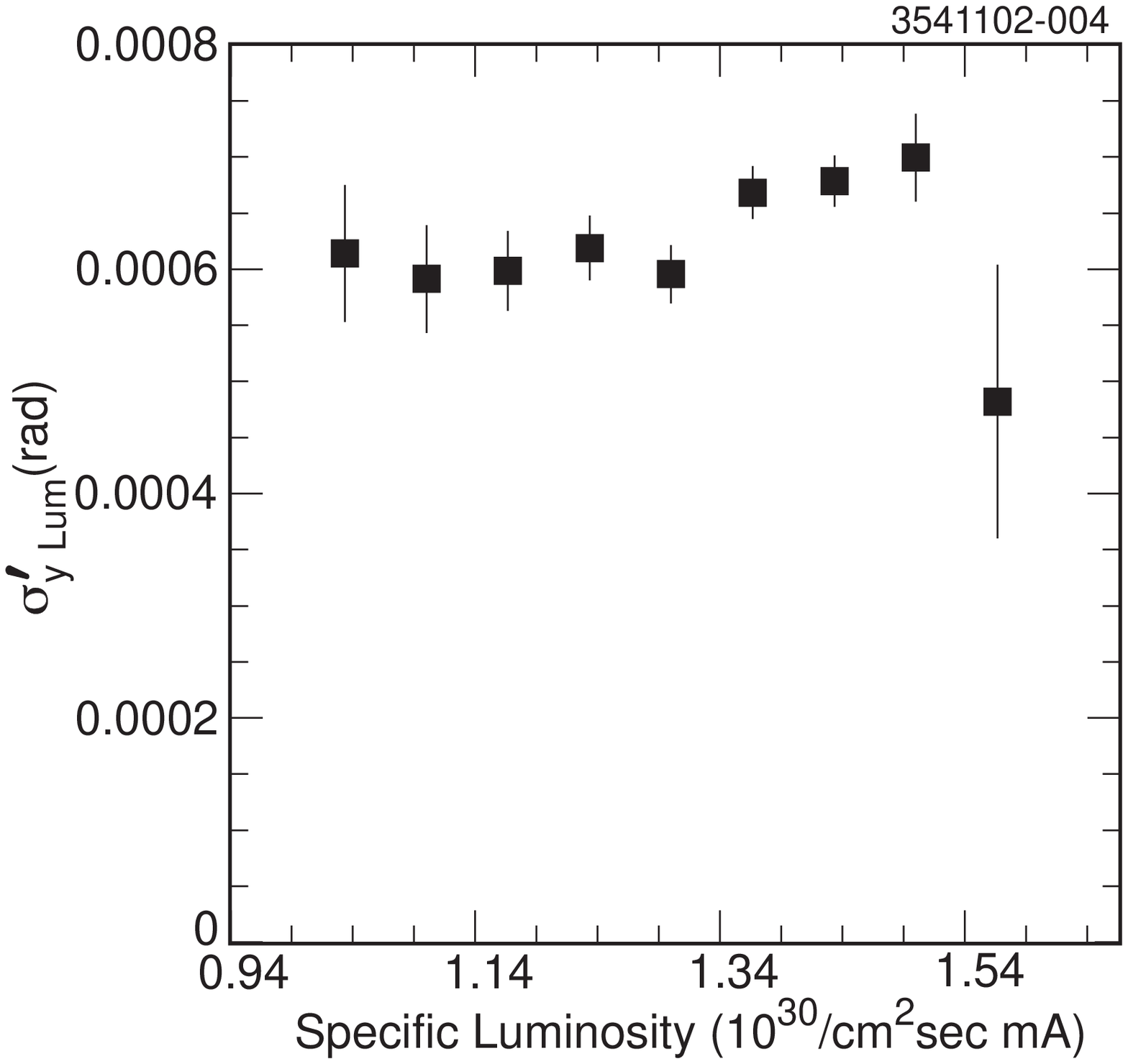,width=6.25in}
\caption{\label{fig:ysigpVlum}The vertical width of the angular 
distribution of the luminous region as a function of specific luminosity.}
\vspace*{2.75in}
\end{center}
\end{figure}

	Figures~\ref{fig:xsigVbmI} 
and~\ref{fig:xsigpVbmI} show the horizontal size and angular spread, 
of the luminous region respectively, as functions of bunch current.
These are after the resolutions discussed above have been taken out.
Note that these are not the beam parameters but the result of the 
colliding beams.  The size is a factor of $\sqrt{2}$ smaller than
the beam size due to the overlap of the two beams producing the luminosity,
and the angular size is a factor of $\sqrt{2}$ larger than the angular
beam spread due to the incoherent collision of the two beams producing the
luminosity.
Figure ~\ref{fig:ysigpVlum} shows the width of the vertical angular
spread of the luminous region as a function of specific luminosity. 
From this data and algebraic manipulation of 
Equations~\ref{eq:shapex},~\ref{eq:shapey},~\ref{eq:angx} and~\ref{eq:angy} 
and the fixed $\beta_y^{\ast}$ discussed above
we extract $\beta_x^{\ast}$, $\epsilon_x$ and $\epsilon_y$. 
 
\begin{figure}
\begin{center}
\epsfig{file=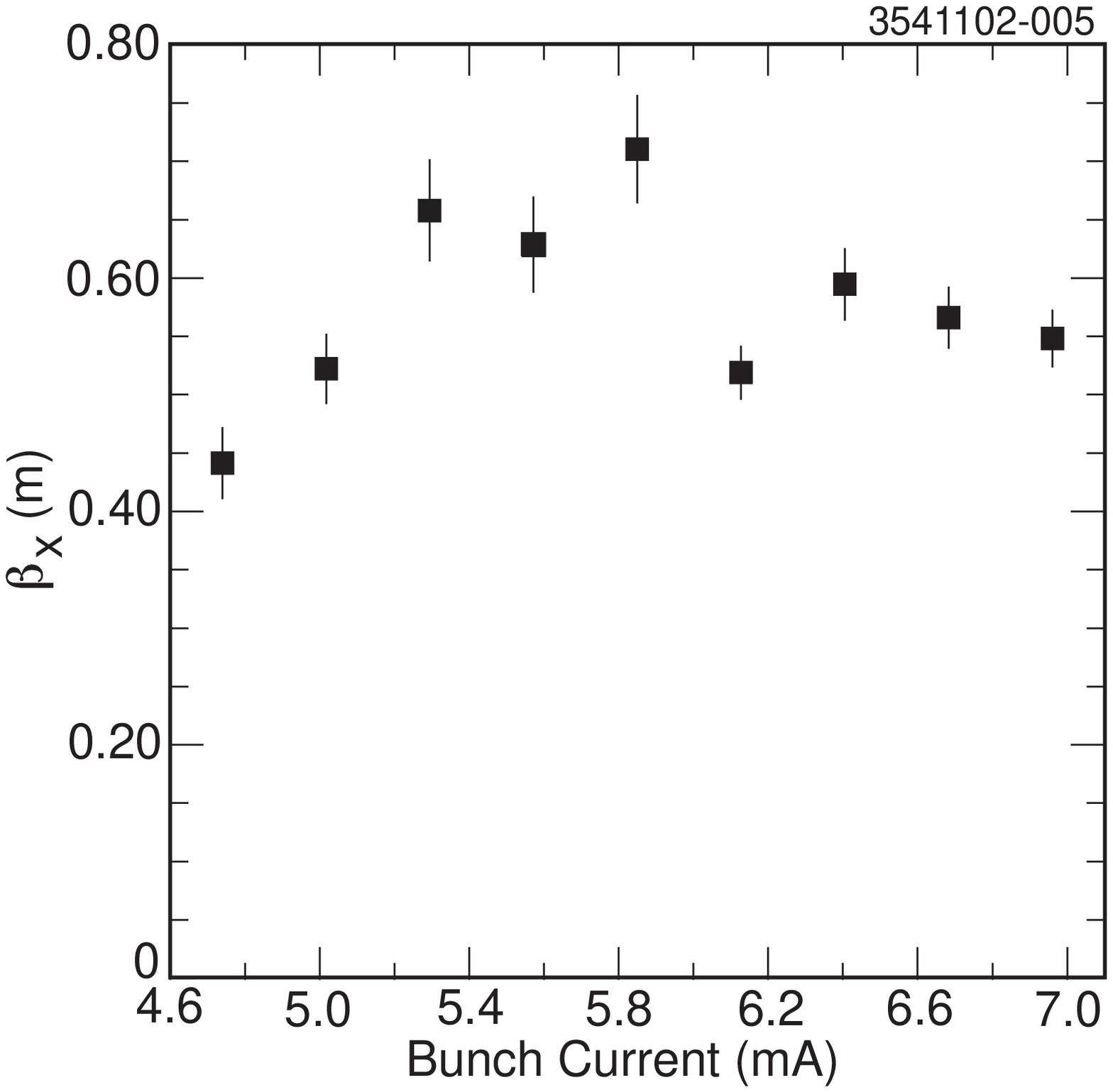,width=6.25in}
\caption{\label{fig:betaxVbmI}The extracted horizontal beta as a function of 
bunch current.}
\vspace*{2.75in}
\end{center}
\end{figure}

\begin{figure}
\begin{center}
\epsfig{file=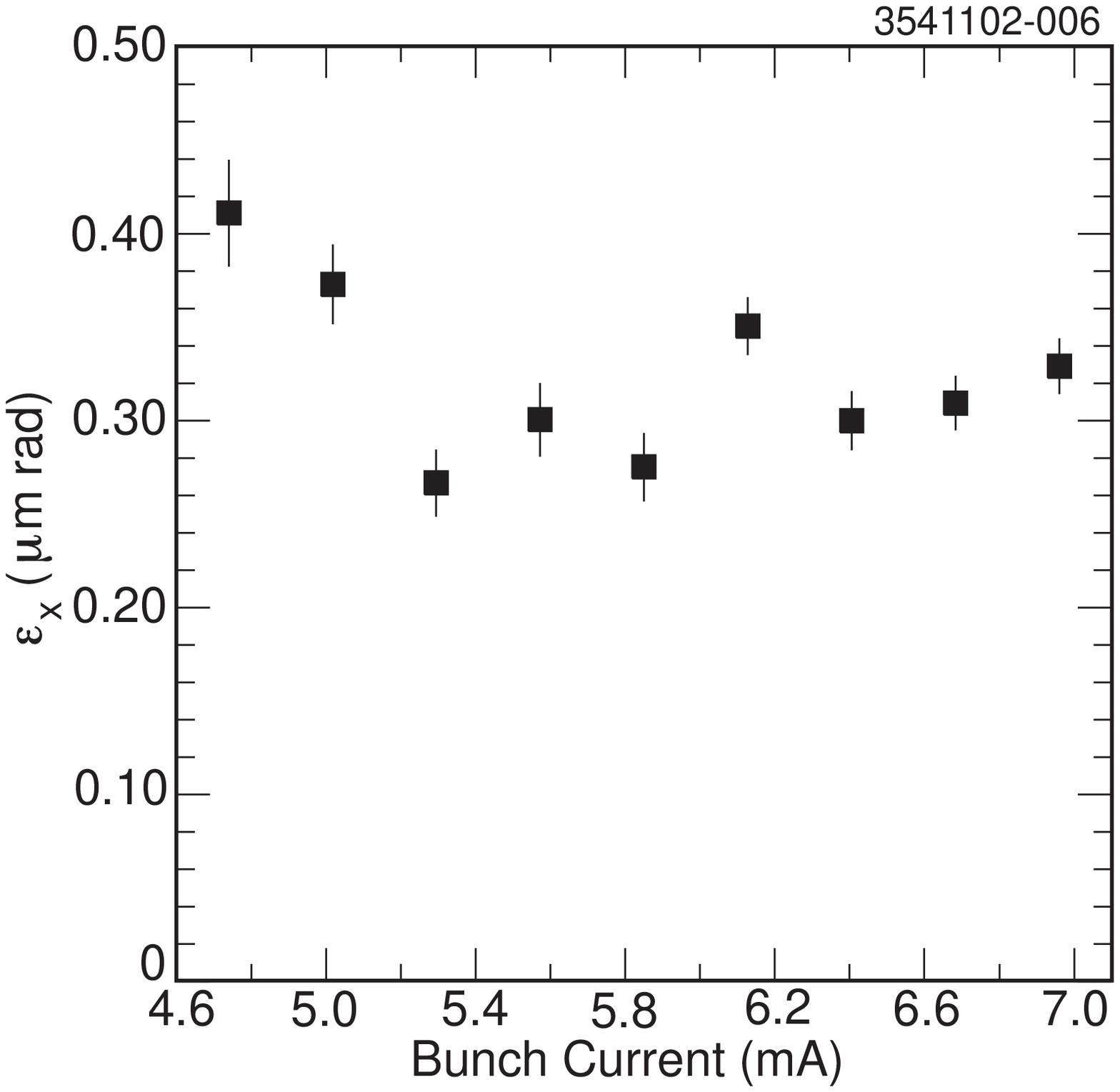,width=6.25in}
\caption{\label{fig:emitxVbmI}The extracted horizontal emittance as a 
function of bunch current.}
\vspace*{2.75in}
\end{center}
\end{figure}

\begin{figure}
\begin{center}
\epsfig{file=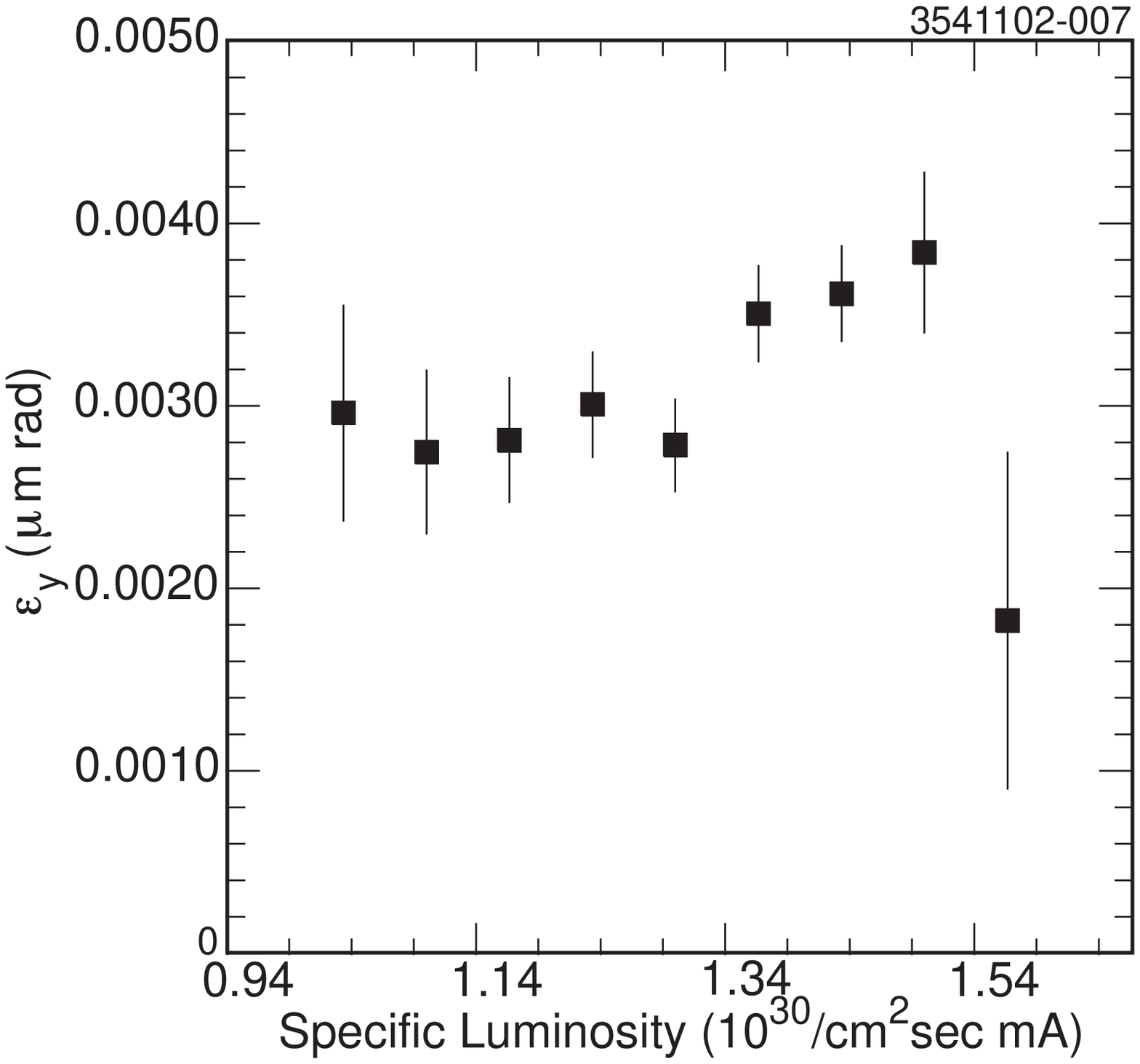,width=6.25in}
\caption{\label{fig:emityVlum}The extracted vertical emittance as a 
function of specific luminosity.}
\vspace*{2.75in}
\end{center}
\end{figure}

The extracted $\beta_x^{\ast}$ is shown in
Figure~\ref{fig:betaxVbmI}.  This result agrees well with our 
expectations based on a previous analysis done which observed the dynamic 
beta effect~\cite{dbeta}.  The expectation was $\beta_x^\ast = 1.2$m reduced 
by roughly a factor of two.  Here we see a $\beta_x^\ast \approx 0.6$m. 
The extracted horizontal emittance as a function of specific luminosity
is displayed in 
Figure~\ref{fig:emitxVbmI}.  We expected an $\epsilon_x = 0.21\mu$m-rad 
while we extracted an $\epsilon_x \approx 0.3\mu$m.  These results again 
agree well with our expectation.  We do not clearly see any dynamic effects.
We also expect an extracted vertical emittance of 
$\epsilon_y = 0.0010\mu$m-rad.  
We extract an $\epsilon_y \approx 0.0030\mu$mrad.
Our extracted value  agrees with an earlier observation of  
$\epsilon_y = (0.0070 \pm 0.0054 \pm 0.0019)\mu$m-rad~\cite{hourglass} and
with values calculated based on the observed luminosity and the best
possible theoretical estimates of the other beam parameters.
This extracted vertical emittance, shown in Figure~\ref{fig:emityVlum} 
appears to 
have dynamic effects.  When a fit to a first order polynomial 
is applied to this data we see a slope that is significant at only
the 2.4 standard deviation level, which we do not claim as
significant.
We observe no visible dynamic effects on vertical emittance although
the data are suggestive.

	These methods can also be applied to observe the beam energy spread
which results in a mismatch of the longitudinal momenta of the two tracks
in these events.  Our resolution is not precise enough to measure
the energy spread, $\Delta E/E$ is expected to be smaller than 0.0007
at CESR, resulting in an upper limit of 0.0015 at 90\% C.L. for this data.

We have developed methods of measuring the emittances
and betas of colliding $e^+e^-$ beams in a non-destructive
manner.  The method depends on a very good understanding
of the resolution of a charged particle tracking system observing
the collision region.
We demonstrate the methods using
$e^{+} e^{-} \to  \mu^{+} \mu^{-}$ events 
from the luminous region of CESR at the
CLEO interaction point. 
We extract the beam
parameters $\beta_x^{\ast}$ and $\epsilon_x$ as functions
of bunch current and 
and $\epsilon_y$ as a function of specific
luminosity.  The observations agree with our expectations
and with measurements made by other methods.
No dynamic effects are observed,
although the data for $\epsilon_y$ are suggestive
of an increase in the emittance with specific luminosity.

\end{document}